\documentclass[aps,twocolumn,pre,floatfix]{revtex4-2}

\usepackage{xcolor,hyperref}
\hypersetup{
   colorlinks,
   linkcolor={blue!50!black},
   citecolor={blue!50!black},
   urlcolor={blue!80!black}
}

\usepackage{epsfig, amsmath}
\usepackage{bm}
\usepackage{soul,ulem}
\usepackage{hyperref}
\usepackage{footmisc}
\usepackage{wrapfig}
\DeclareMathAlphabet{\mathitbf}{OML}{cmm}{b}{it}

\newcommand{\zerovector}{\bm{0}}

\renewcommand{\=}{\!=\!}

\newcommand{\xv}{\mathitbf x}
\newcommand{\zv}{\mathitbf z}
\newcommand{\phiv}{\bm{\phi}}
\newcommand{\piv}{\bm{\pi}}
\newcommand{\psiv}{\bm{\psi}}

\newcommand{\calBold}[1]{\mbox{\boldmath${\cal #1}$}}
\newcommand{\dbar}{{\,\mathchar'26\mkern-12mu d}}

\newcommand{\wbp}{\omega_{\mbox{\tiny BP}}}

\definecolor{darkGreen}{RGB}{0,100,0}

\begin{document}

\title{Boson-peak vibrational modes in glasses feature\\ hybridized phononic and quasilocalized excitations}
\author{Edan Lerner$^{1}$}
\email{e.lerner@uva.nl}
\author{Eran Bouchbinder$^{2}$}
\email{eran.bouchbinder@weizmann.ac.il}
\affiliation{$^{1}$Institute of Theoretical Physics, University of Amsterdam, Science Park 904, 1098 XH Amsterdam, the Netherlands\\
$^{2}$Chemical and Biological Physics Department, Weizmann Institute of Science, Rehovot 7610001, Israel}

\begin{abstract}
A hallmark of structural glasses and other disordered solids is the emergence of excess low-frequency vibrations, on top of the Debye spectrum $D_{\mbox{\tiny Debye}}(\omega)$ of phonons ($\omega$ denotes the vibrational frequency), which exist in any solid whose Hamiltonian is translationally invariant. These excess vibrations --- a signature of which is a THz peak in the reduced density of states $D(\omega)/D_{\mbox{\tiny Debye}}(\omega)$, known as the boson peak --- have resisted a complete theoretical understanding for decades. Here, we provide direct numerical evidence that vibrations near the boson peak consist of hybridizations of phonons with many \emph{quasilocalized excitations}, the latter were recently shown to generically populate the low-frequency tail of the vibrational spectra of structural glasses quenched from a melt and of disordered crystals. Our results suggest that quasilocalized excitations exist up to and in the vicinity of the boson-peak frequency, and hence constitute fundamental building blocks of the excess vibrational modes in glasses.   
\end{abstract}

\maketitle

\section{introduction}
\label{sec:introduction}

The behavior of many mechanical, transport and thermodynamic properties of solids can be explained based on their vibrational spectra~\cite{kittel2005introduction}. Examples include the specific heat~\cite{kittel2005introduction}, heat transport~\cite{vincenzo_transport_prl_2009}, elastic moduli~\cite{lutsko,exist}, and nonlinear failure mechanisms~\cite{david_collaboration_2020,david_fracture_mrs_2021}. Unlike crystalline materials, for which the low-frequency spectrum is well-understood~\cite{kittel2005introduction}, there are still open questions regarding the nature of low-frequency vibrational excitations in disordered solids~\cite{soft_potential_model_1991,Schober_Oligschleger_numerics_PRB,Gurevich2003,Gurevich2007,Lubchenko1515,matthieu_PRE_2005,tanaka_boson_peak_2008,procaccia_boson_peak_prb_2009,Chumakov_2011_bosonPeak,sokolov_boson_peak_scale,Schirmacher_2013_boson_peak,eric_boson_peak_emt,silvio,JCP_Perspective,tanaka_2d_modes_2022,disordered_crystals_prl_2022}. In particular, disordered solids feature excess low-frequency modes on top of the Debye phononic spectrum $D_{\mbox{\tiny Debye}}(\omega)\!\sim\!\omega^{\dbar-1}$ (here $\omega$ denotes the vibrational frequency and $\dbar$ is the dimension of space), which stem from their intrinsic structural disorder and micromechanical frustration~\cite{shlomo}.

Observing the abundance of excess low-frequency excitations in disordered solids is commonly achieved, especially experimentally, by dividing the measured vibrational density of states (VDoS) $D(\omega)$ by the expectation $D_{\mbox{\tiny Debye}}(\omega)$ based on Debye's theory of low-frequency phonons. The reduced VDoS $D(\omega)/D_{\mbox{\tiny Debye}}(\omega)$ generically exhibits a peak --- commonly referred to as the \emph{boson peak} --- at a characteristic frequency commonly referred to as the \emph{boson-peak frequency} $\wbp$, usually in the THz range~\cite{baldi2010sound,sokolov_boson_peak_scale}.

Several explanations for the emergence of the boson peak have been suggested in the literature in the past decades. These include effects of the unjamming criticality~\cite{mw_EM_epl,eric_boson_peak_emt,new_variational_argument_epl_2016}, a broadening of the so-called van-Hove singularity~\cite{Chumakov_2011_bosonPeak}, effects of elastic-moduli spatial fluctuations~\cite{Duval_prb_1998, Schirmacher_2013_boson_peak}, and others~\cite{taraskin_lattice_disorder_prl_2001,Lubchenko1515,Parisi_boson_peak_2003,tanaka_boson_peak_2008,procaccia_boson_peak_prb_2009}. We highlight here in particular Refs.~\cite{Schober_Oligschleger_numerics_PRB,Gurevich2003,Gurevich2007}, where it was suggested that quasilocalized excitations (QLEs) play a crucial role, not just in the low-frequency $(\omega\!\to\!0)$ tail of the VDoS, but also in the physics of the boson peak at higher frequencies.

In recent years, computational studies have established many of the generic properies of QLEs; in particular, it has been shown that QLEs populate the low-frequency tail of structural glasses~\cite{modes_prl_2016}, independently of the details of the interparticle potential~\cite{modes_prl_2020} or the glass formation protocol~\cite{pinching_pnas,LB_modes_2019}. In a particular range of glass-sample sizes~\cite{phonon_widths}, QLEs assume the form of (quasilocalized) vibrational modes. In addition, it has been demonstrated that these low-frequency excitations follow a universal nonphononic VDoS, scaling as $D_{\mbox{\tiny QLE}}(\omega)\!\sim\!\omega^4$~\cite{JCP_Perspective} in any spatial dimension $\dbar\!\ge\!2$~\cite{modes_prl_2018}. Finally, the quasilocalized nature of QLEs has been elucidated as they were shown to feature a disordered core of linear size $\xi_{\rm g}$ (of a few atomic distances) accompanied by long-range displacement fields that decay as $1/r^{\dbar-1}$~\cite{JCP_Perspective}. A review of past and recent efforts to understand the emergent statistics and properties of QLEs can be found in~\cite{JCP_Perspective}. 

However, possible relations between QLEs and vibrations at higher frequencies, especially at and in the vicinity of the boson-peak frequency, have not yet been fully elucidated. In fact, in Ref.~\cite{tanaka_2d_modes_2022}, it was argued that vibrations around $\wbp$ --- coined `stringlets' therein --- are distinct from QLEs (termed `four-leaf-like' vibrations in~\cite{tanaka_2d_modes_2022}, due to their quadrupolar far fields), the latter presumably only emerge at lower frequencies $\omega\!\ll\!\wbp$.

Here, we employ numerical simulations of a two-dimensional (2D) glass former (see model details in Sect.~\ref{sec:model}) and a variety of recently developed tools to demonstrate that harmonic vibrations at the vicinity of the boson-peak frequency $\wbp$ consist of hybridization (mixing) of QLEs and phonons. In particular, this is achieved by utilizing the framework put forward in Ref.~\cite{pseudo_harmonic_prl}, which allows to effectively de-hybridize individual QLEs from hybridized vibrations. Our findings support the assertion of Refs.~\cite{Schober_Oligschleger_numerics_PRB,Gurevich2003,Gurevich2007} that QLEs are the origin of the boson peak, and indicate that no other classes of excitations --- distinct from QLEs --- emerge in the vicinity of the boson-peak frequency, e.g.~as claimed recently~\cite{tanaka_2d_modes_2022}. These results further underline the importance of obtaining a complete understanding of the generic emergence of QLEs in disordered solids.

This work is structured as follows; in Sect.~\ref{sec:model}, we describe the glass-forming model employed. In Sect.~\ref{sec:qlm_detection}, we explain how QLEs can be detected and how their properties are measured using the methodological framework of Ref.~\cite{pseudo_harmonic_prl}. Section~\ref{sec:superposition} describes how to construct superpositions of many QLEs that resemble the nonphononic parts of vibrational modes at the vicinity of the boson-peak frequency. In Sect.~\ref{sec:results}, the main results are presented, followed by an analysis of and discussion about the `excess-modes' on top of the Debye phononic VDoS in Sect.~\ref{sec:excess_modes}. Sect.~\ref{sec:discussion} provides a summary of this work and some future research questions. 

\vspace{-0.2cm}

\section{glass-forming model}
\label{sec:model}

We employed the same well-studied two-dimensional ($\dbar\=2$) glass forming model of Ref.~\cite{tanaka_2d_modes_2022}. This model is composed of a 50:50 binary mixture of `large' and `small' particles of equal mass, interacting via a purely repulsive $\sim\!r^{-10}$ pairwise potential ($r$ denotes the pairwise distance between particles) that is smoothed up to two derivatives at some cutoff pairwise distance. The size ratio between the effective diameters of the `large' and `small' particles is 1.4, and the number density is set to $N/V\!=\!0.86$ (in simulation units). Further details about the interaction potential and the units employed can be found, e.g., in Ref.~\cite{cge_paper}. 

We prepared 1000 independent glassy samples of $N\!=\!8100$ particles, using the aforementioned glass-forming model. This ensemble size is required to ensure the statistical convergence of the VDoS. The computer glass transition temperature of this model is estimated as $T\!=\!0.5$ (in simulation units). We prepared glasses by first equilibrating the system at high-temperature liquid states with $T\!=\!1$ (again in simulation units). Then, we quenched the system at a finite cooling rate of $\dot{T}\!=\!10^{-3}$, until reaching $T\!=\!0.1$. We then created a zero-temperature glass by removing the excess heat via a standard conjugate-gradient minimization algorithm. 
For the sake of reproducibility, we report here the values of some basic observables: the mean shear and bulk elastic moduli of our glass ensemble are $G\!\approx\!18.3$ and $K\!\approx\!122.1$, respectively, reported in simulational units. The resulting shear wave speed, sound wave speed, and Debye frequency are $c_{\rm s}\!\approx\!4.6$, $c_{\ell}\!\approx\!12.78$ and $\omega_{\rm D}\!=\left( \frac{ 8 \pi (N/V) }{c_{\rm s}^{-2} + c_\ell^{-2} } \right)^{1/2}\!\approx\!20.1$, respectively. The `disorder parameter' $\chi\!\equiv \!\sqrt{N}\,\mbox{std}(G)/\mbox{mean}(G)\!\approx\!3.8$ (measured over our \emph{ensemble} of glassy samples as explained in~Ref.~\cite{sticky_spheres1_karina_pre2021}). The mean energy-per-particle is $u/N\!\approx\!3.87$ and the mean hydrostatic pressure is $19.36$, again reported in simulational units.

\begin{figure*}[ht!]
  \includegraphics[width = 1\textwidth]{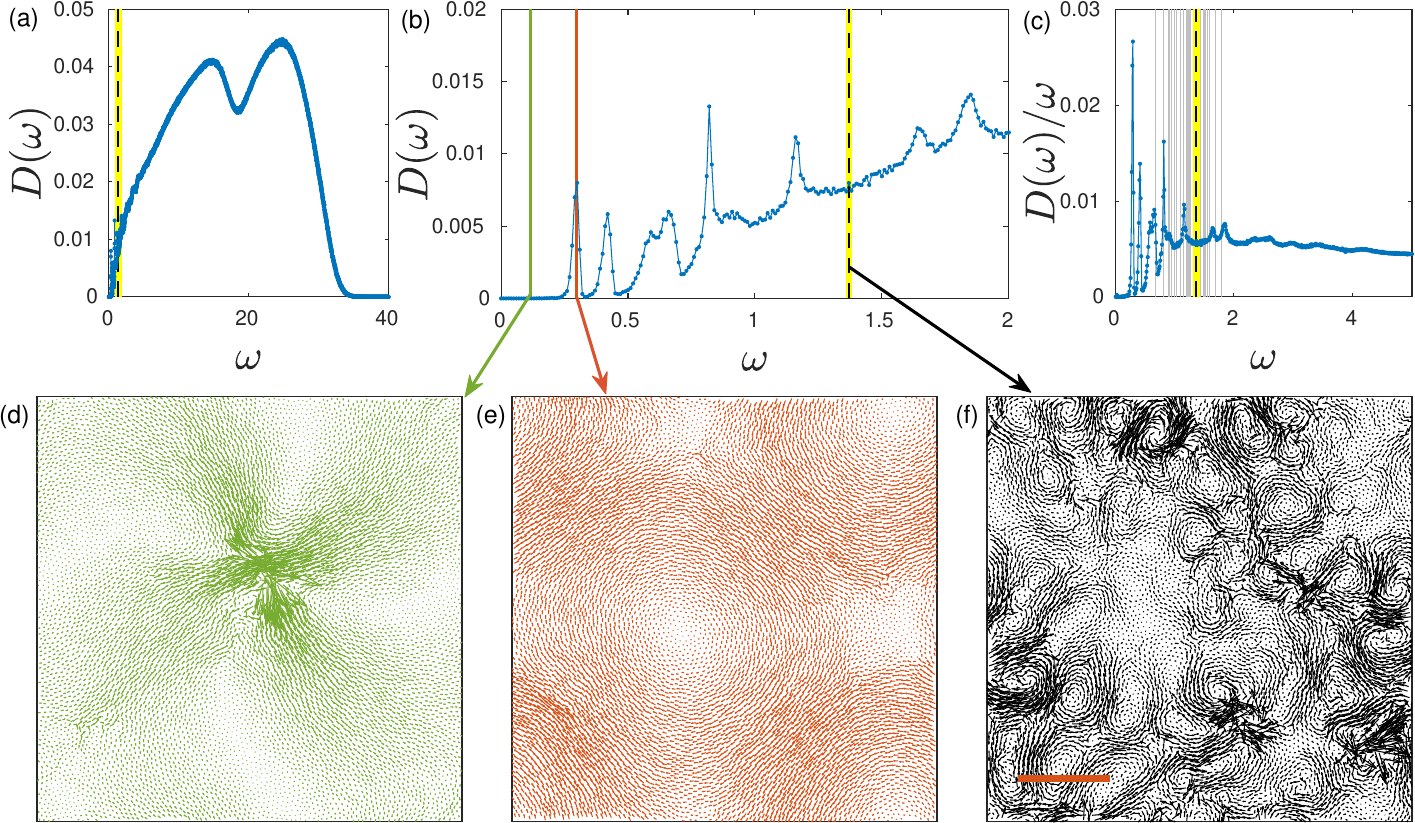}
  \caption{\footnotesize (a) The vibrational spectrum $D(\omega)$ of the 2D computer glass former employed in this work. The vertical dashed line marks the boson-peak frequency (defined as shown in panel (c)). (b) A zoomed-in view on the low-frequency regime of the vibrational spectrum shown in panel (a). The vertical green and red lines mark the frequencies of the quasilocalized vibrational mode shown in panel (d), and of the phononic vibrational mode shown in panel (e), respectively. The vertical dashed line marks the boson-peak frequency (cf.~panel (a)), populated by vibrational modes of similar spatial features as the one shown in panel (f). Modes are plotted by quivering the eigenfunctions obtained from diagonalizing the Hessian matrix. The red bar in panel (f) represents the phononic wavelength $2\pi c_{\rm s}/\wbp\!\approx\! 21.1$, which matches well the mode's wave-like background patterns, and see further examples in Fig.~\ref{fig:fig5} below. (c) The reduced spectrum $D(\omega)/\omega$ is plotted vs.~$\omega$; following~\cite{tanaka_2d_modes_2022}, we estimate the maximum to occur at $\wbp\!\approx\!1.37$ (vertical dashed line). The faded vertical lines represent the frequencies of the QLEs used in the analysis of Sect.~\ref{sec:results}, see text for additional details.}
  \label{fig:fig1}
\end{figure*}

\subsection{Vibrational spectrum}

In Fig.~\ref{fig:fig1}, we present key features of the vibrational spectrum of the employed glass-forming model. The full VDoS, obtained by diagonalizing the Hessian matrix (see definition below), is shown in Fig.~\ref{fig:fig1}a; the dashed vertical line (also in Figs.~\ref{fig:fig1}b-c) marks the boson-peak frequency, estimated following~\cite{tanaka_2d_modes_2022} as $\wbp\!\approx\!1.37$. The latter is defined as the maximum of the reduced VDoS $D(\omega)/\omega\!\sim\!D(\omega)/D_{\mbox{\tiny Debye}}(\omega)$, shown in Fig.~\ref{fig:fig1}c. Figure~\ref{fig:fig1}b presents a zoomed-in view of the low-frequency regime of the VDoS of Fig.~\ref{fig:fig1}a. The distinct peaks in Fig.~\ref{fig:fig1}b correspond to discrete phonon bands, broadened by disorder~\cite{phonon_widths}. Vibrational modes occurring below the lowest-frequency phonon band are typically quasilocalized modes~\cite{JCP_Perspective}; an example of one with $\omega\!\approx\!0.11$ is shown in Fig.~\ref{fig:fig1}d. Panels (e) and (f) show examples of a long-wavelength phonon and a boson-peak mode, respectively. While the boson-peak mode (i.e.~a mode whose vibrational frequency is close to $\wbp$) in panel (f) is clearly complex, it features a distinct phononic background pattern. To highlight this point, we added to panel (f) a red bar, which corresponds to the wavelength $2\pi c_{\rm s}/\wbp\!\approx\!21.1$ (in simulational units) of a plane-wave phonon of frequency $\wbp$. The red bar matches well the spatial periodicity of the wave-like pattern seen in the mode's structure. Similar comparisons are shown in Fig.~\ref{fig:fig3}e and in Appendix~\ref{sec:AppendixA}.

\section{Definition and calculation of quasilocalized excitations}
\label{sec:qlm_detection}

Previous work on soft excitations in glasses~\cite{SciPost2016, episode_1_2020,pseudo_harmonic_prl} has demonstrated that QLEs can be isolated and dehybridized from other QLEs and phononic excitations using various cost functions, which make use of the known spatial localization properties of QLEs. The arguments of these cost functions are generalized directions $\zv$ in the glass's $N\dbar$-dimensional configuration space. 

\begin{figure*}[ht!]
  \includegraphics[width = 1\textwidth]{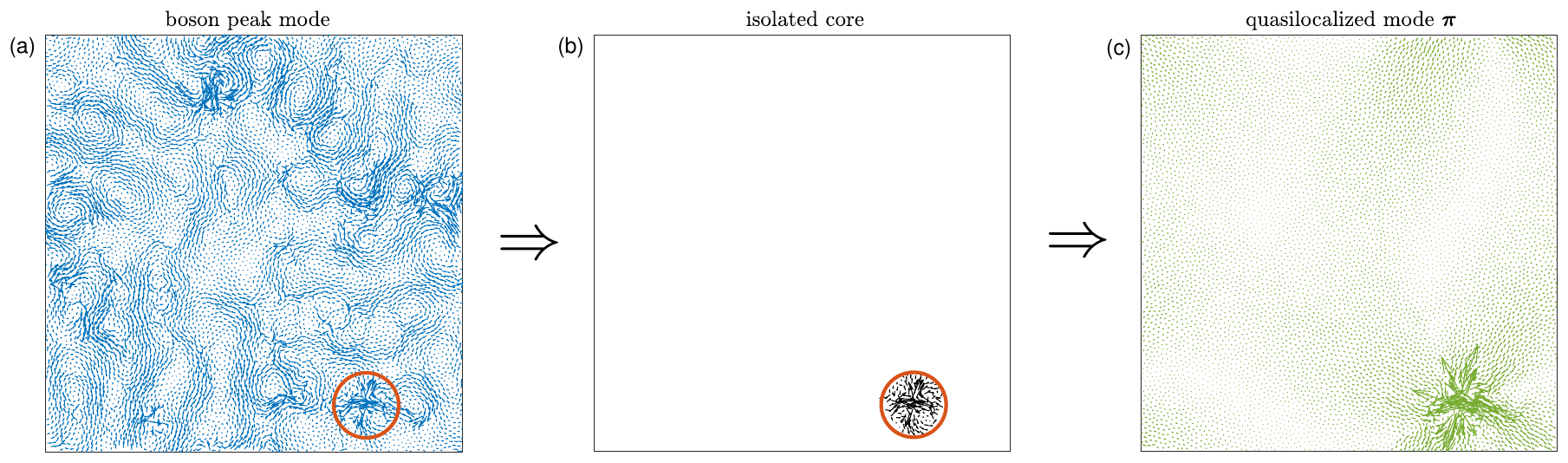}
  \caption{\footnotesize Illustration of the scheme employed for extracting QLEs in computer glasses. (a) A boson-peak vibrational mode. The circle marks a region identified as a QLE-core, featuring a quadrupolar spatial structure. (b) Given the QLE-core, the rest of the field is suppressed. The resulting isolated core is used as the initial conditions for the minimization of the cost function ${\cal C}(\zv)$ (cf.~Eq.~(\ref{eq:phm_cost_function})). (c) The QLE resulting from the cost function minimization, featuring a core that is very similar to the one shown in panel (b) and long-range algebraic tails. }
  \label{fig:fig2}
\end{figure*}

In what follows, we employ a cost function ${\cal C}(\zv)$ that only depends on the harmonic approximation to the potential energy of the glass~\cite{pseudo_harmonic_prl}. It reads
\begin{equation}\label{eq:phm_cost_function}
    {\cal C}(\zv) \equiv \frac{\zv\cdot\calBold{H}\cdot\zv}{\sum_{\langle ij\rangle}\zv_{ij}\cdot\zv_{ij}}\,,
\end{equation}
where $\calBold{H}\!\equiv\!\frac{\partial^2U}{\partial\xv\partial\xv}$ is the Hessian matrix of the potential energy $U(\xv)$, $\xv$ are the particles' coordinates, and $\langle ij \rangle$ label a pair of interacting particles. Local minima of $C(\zv)$ occur at directions $\zv\=\piv$, which were termed pseudo-harmonic modes in Ref.~\cite{pseudo_harmonic_prl}; since ${\cal C}(\zv)$ assumes local minima at $\piv$'s, the latter satisfy
\begin{equation}
    \frac{\partial {\cal C}}{\partial\zv}\bigg|_{\piv} = \zerovector\,.
\end{equation}
Here, we obtain solutions $\piv$ to the above equation by minimizing ${\cal C}(\zv)$ using a standard conjugate-gradient algorithm. An open-source package to carry out such calculations is provided in Ref.~\cite{package}. Example QLEs detected using this procedure are shown in Figs.~\ref{fig:fig2}c and~\ref{fig:fig6}.

The initial conditions for the minimizations described above are taken to be isolated QLE-cores that are highly nonaffine, which we identify as spatial quadrupolar structures in boson-peak vibrational modes, as illustrated in Fig.~\ref{fig:fig2}. During the calculation, we record new QLEs if their spatial overlap with previously detected QLEs falls below 0.99. The detected QLEs are then used as a ``library/catalog'' in the analysis to be described next.

\begin{figure*}[ht!]
  \includegraphics[width = 0.85\textwidth]{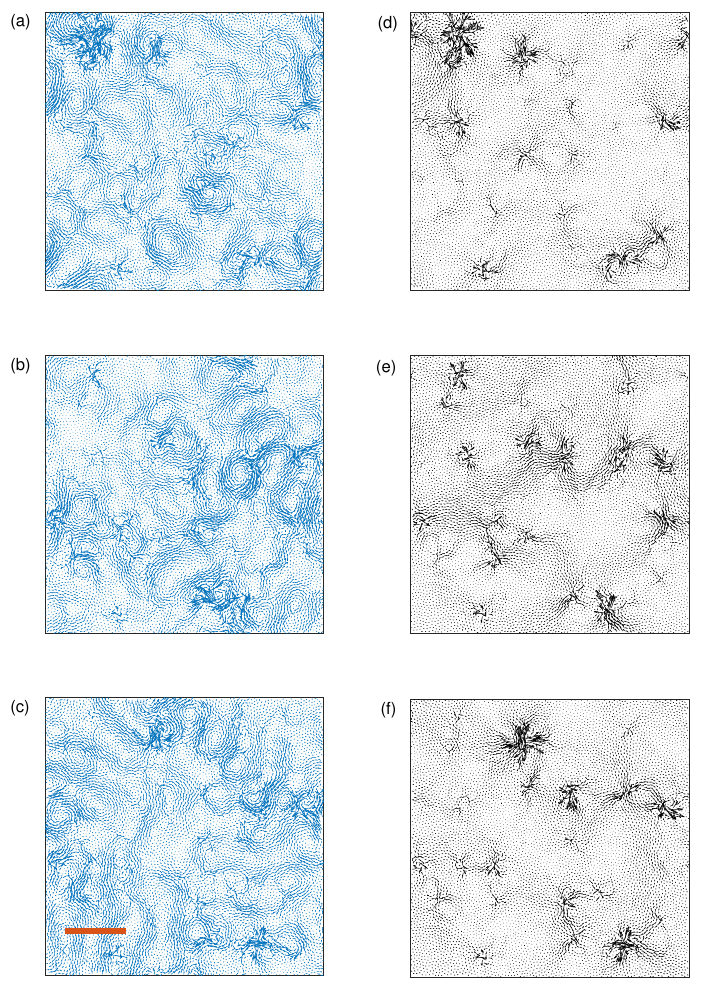}
  \caption{\footnotesize Comparison between vibrational boson-peak modes and the corresponding constructed superpositions of quasilocalized modes. Panels (a)-(c) show vibrational modes $\psiv$, whose frequencies $\omega\!\approx\!\wbp$. The red bar shows the expected phononic wavelength, obatined from the plane-wave relation $2\pi c_{\rm s}/\wbp\!\approx\!21.1$ (in simulational units) at the boson peak frequency; it matches well the characteristic scale of the wave-like background patterns seen in these boson-peak modes, and see additional examples in Fig.~\ref{fig:fig5} below. Panels (d)-(f) show the modes $\phiv_\dagger$ constructed following the scheme spelled out in Sect.~\ref{sec:superposition} --- solely from the $31$ QLEs detected as explained in Sect.~\ref{sec:qlm_detection}. Our scheme cannot reproduce the short wavelength phononic background, but captures well the nonaffine parts of the vibrational modes. The values of the overlaps $\psiv\!\cdot\!\phiv_\dagger$ between the vibrational modes $\psiv$ and the constructed fields $\phiv_\dagger$ are 0.49, 0.42 and 0.41 for the pairs of fields in the upper, middle and bottom rows, respectively.}
  \label{fig:fig3}
\end{figure*}

\section{Constructing superpositions of quasilocalized modes}
\label{sec:superposition}

Given a harmonic vibrational mode $\psiv$, i.e.~an eigenmode of the Hessian matrix $\calBold{H}$, we search for a normalized, linear superposition $\phiv$ of QLEs of the form
\begin{equation}
    \phiv = \sum_\ell c_\ell\,\piv_\ell\,,
\end{equation}
which has the maximal absolute value of the overlap $\phiv\cdot\psiv$. To this aim, we construct another cost function ${\cal F}(\{c_\ell\})$ that depends on the coefficients $c_\ell$, which reads
\begin{equation}
    {\cal F}(\{c_\ell\}) = \frac{(\phiv\cdot\psiv)^2}{(\phiv\cdot\phiv)(\psiv\cdot\psiv)}\,.
\end{equation}
By maximizing this cost function with respect to the coefficients $c_\ell$, we obtain the superposition $\phiv_\dagger$ that has the largest overlap with the vibrational mode $\psiv$. In what follows, we study the properties of the QLE-constructed modes $\phiv_\dagger$ and their similarity (or lack thereof) to vibrational modes at the vicinity of the boson-peak frequency $\wbp$.

\section{results}
\label{sec:results}

Our main goal is to assess whether QLEs play the role of fundamental building blocks of vibrational modes in the vicinity of the boson-peak frequency $\wbp$. To this aim, we show in the top row of Fig.~\ref{fig:fig3} three vibrational modes whose frequencies dwell in the vicinity of the boson-peak frequency $\wbp$. The expected phononic wavelength, assuming the plane-wave relation $2\pi c_{\rm s}/\wbp\!\approx\! 21.1$, is represented by the red bar in Fig.~\ref{fig:fig3}c. We use these three boson-peak vibrational modes to detect 31 QLEs as described in Sect.~\ref{sec:qlm_detection}. The frequencies of the detected modes~\cite{footnote} are represented by the faded vertical lines shown in Fig.~\ref{fig:fig1}c; they clearly dwell at and in the vicinity of the boson-peak frequency $\wbp\!\approx\!1.37$. The highest and lowest frequency QLEs detected are shown in Appendix~\ref{sec:AppendixB}. We next use the extracted set of QLEs (``library/catalog'') to construct the fields $\phiv_\dagger$ following the procedure explained in Sect.~\ref{sec:superposition}; the constructed fields $\phiv_\dagger$ are shown in the bottom row of Fig.~\ref{fig:fig3}. 

As expected, the superpositions $\phiv_\dagger$ are unable to generate short-wavelength phononic patterns as seen in the spatial structure of the boson-peak vibrations of Figs.~\ref{fig:fig3}a-c. However, the disordered, highly nonaffine cores in $\phiv_\dagger$ remarkably match the same patterns as seen in the harmonic vibrational modes. Since our QLE-detection scheme is not fully exhaustive, some highly nonaffine parts of the vibrational modes are absent in the reconstructed fields. Nevertheless, our results clearly indicate that vibrational modes near the boson-peak frequency consist of hybridized phonons and QLEs (see also a comparison to lower frequencies, presented in Appendix~\ref{sec:AppendixA}). This is the main result of this contribution. It is important to note that the boson peak modes analyzed in Fig.~\ref{fig:fig3} were extracted from a randomly selected glass sample, highlighting the generic nature of our main result.

\section{Excess vibrational modes}
\label{sec:excess_modes}

\begin{figure}[ht!]
  \includegraphics[width = 0.5\textwidth]{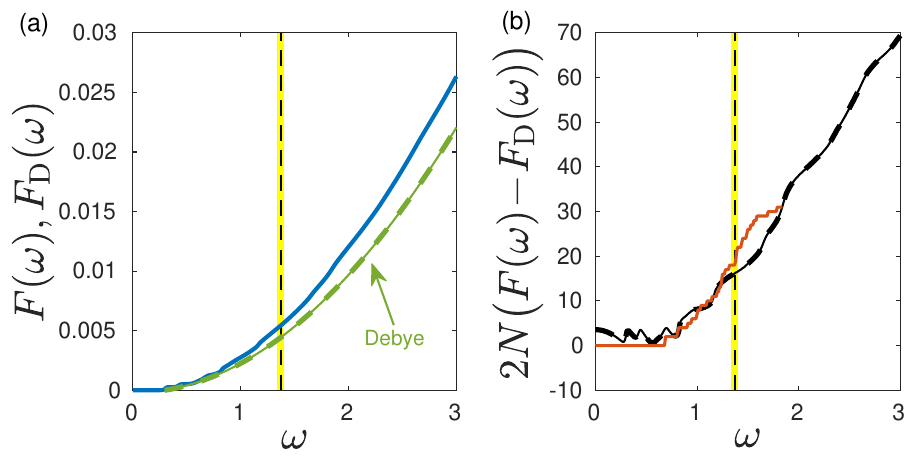}
  \caption{\footnotesize (a) The cumulative VDoS $F(\omega)\!\equiv\!\int_0^\omega D(\omega')d\omega'$ (continuous blue line) extracted from our glass ensemble and the cumulative phononic (Debye) VDoS $F_{\rm D}(\omega)$ (dashed green line) given in Eq.~(\ref{eq:cumulative_phonons}). At low frequencies, the two curves overlap, but start to deviate from each other before $\wbp$ (marked by the vertical highlighted lines in both panels), indicating the emergence of excess (nonphononic) modes. (b) The difference $F(\omega)\!-\!F_{\rm D}(\omega)$ multiplied by $2N$, which quantifies the number of excess modes up to frequency $\omega$ in a single glass sample, is plotted in the dashed black line. The thin red line represents the number of QLEs, up to frequency $\omega$, obtained using the QLE-detection method applied as in Fig.~\ref{fig:fig3} above, see text for further discussion. The two curves are in good quantitative agreement with each other.}
  \label{fig:fig4}
\end{figure}

While our QLE-detection scheme may not be entirely exhaustive, we can assess how the number of excess modes as gleaned from the bare VDoS of our computer glasses compares with the number of QLEs we have detected. To this aim, we plot in Fig.~\ref{fig:fig4}a the cumulative VDoS $F(\omega)\!\equiv\!\int_0^\omega D(\omega')d\omega'$ of our computer glass ensemble. Next, we invoke Debye's theory of phonons~\cite{kittel2005introduction} (which is a continuum approximation) and superimpose in Fig.~\ref{fig:fig4}a the cumulative phononic density of states $F_{\rm D}(\omega)$ up to frequency $\omega$ in a system of linear size $L$, namely 
\begin{equation}\label{eq:cumulative_phonons}
F_{\rm D}(\omega) = \int\limits_{2\pi c_{\rm s}/L}^\omega \frac{2\omega'}{\omega_{\rm D}^2}d\omega' = \frac{\omega^2-(2\pi c_{\rm s}/L)^2}{\omega_{\rm D}^2}\,.
\end{equation}
The latter is plotted as the dashed line in Fig.~\ref{fig:fig4}a (recall that our glasses are two-dimensional), and note that the lower integration limit in Eq.~\eqref{eq:cumulative_phonons} is the minimal phonon frequency in a system of linear size $L$. At low frequencies, the two curves overlap, and before $\wbp$ (marked by the highlighted vertical dashed lines in Fig.~\ref{fig:fig4}) the curves depart from each other, indicating the emergence of excess vibrational modes on top of phonons.

In Fig.~\ref{fig:fig4}b, we show (dashed black line) the difference $F(\omega)\!-\!F_{\rm D}(\omega)$~\cite{moriel2023boson}, multiplied by $2N$, which quantifies the \emph{number of excess modes} up to frequency $\omega$ in a single glass sample. We superimpose on it (thin red line) the number of QLEs, up to frequency $\omega$, obtained using the QLE-detection method applied as in Fig.~\ref{fig:fig3} above. The two curves reveal good quantitative agreement. In particular, note that our QLE-library contains modes with frequencies up to about $\omega\!\approx\!1.8$ (see the highest frequency QLE of our library in Fig.~\ref{fig:fig6}b below). On average, the expected number of excess modes up to $\omega\!=\!1.8$ is $\approx\!28$, according to the data of Fig.~\ref{fig:fig4}b, i.e.~in good agreement with the actual number of QLEs (31) in our library. This agreement lends quantitative support to our computational approach and conclusions.

\newpage

\section{Summary and outlook}
\label{sec:discussion}

Using recently developed tools to single out QLEs in model glass formers~\cite{SciPost2016,pseudo_harmonic_prl,episode_1_2020}, we provided strong visual evidence that vibrational modes at the vicinity of the boson-peak frequency $\wbp$ in a generic 2D glass former consist of hybridized QLEs and phonons. Our results suggest that the `excess modes' universally seen as a boson-peak in glasses' reduced spectra (and in disordered crystals~\cite{disordered_crystals_prl_2022}) are intimately related to the emergence of QLEs in those systems. In addition, we establish that quasilocalized excitations are plentiful (i.e.~are not rare) --- and coupled among themselves and with phonons --- at and in the vicinity of the boson-peak frequency, consistent with the suggestions of~\cite{Schober_Oligschleger_numerics_PRB,Gurevich2003,Gurevich2007}, but at odds with recent claims~\cite{tanaka_2d_modes_2022}. 

Our findings reinforce the statement of Ref.~\cite{experimental_local_vibrations_boson_peak_prl_2006} --- based on experimental evidence --- that vibrations constituting the boson peak in amorphous solids have a local character. In addition, the construction of boson-peak vibrational modes by hybridized QLEs (and phonons) somewhat echoes the ideas of Ref.~\cite{new_variational_argument_epl_2016}, where it was suggested that excess vibrational modes can be constructed from (quasilocalized) responses to local force dipoles.

Our results, which indicate that QLEs play important roles in the physics of the boson peak in glasses, may also imply that the boson-peak frequency $\wbp$ is related to the characteristic frequency $\omega_{\rm g}$ of QLEs~\cite{cge_paper,pinching_pnas,JCP_Perspective}. The latter can be identified from the QLEs VDoS, whose dimensional form reads $D_{\mbox{\tiny QLE}}(\omega)\!\sim\!\omega^4/\omega_{\rm g}^5$, as explained in~\cite{pinching_pnas,JCP_Perspective}. Indeed, a close relation between $\omega_{\rm g}$ and $\wbp$ has been suggested and indirectly supported in~\cite{pinching_pnas}. The findings of the present work support such an intimate relation, which should be further explored in future work. 

\begin{figure*}[ht!]
  \includegraphics[width = 1\textwidth]{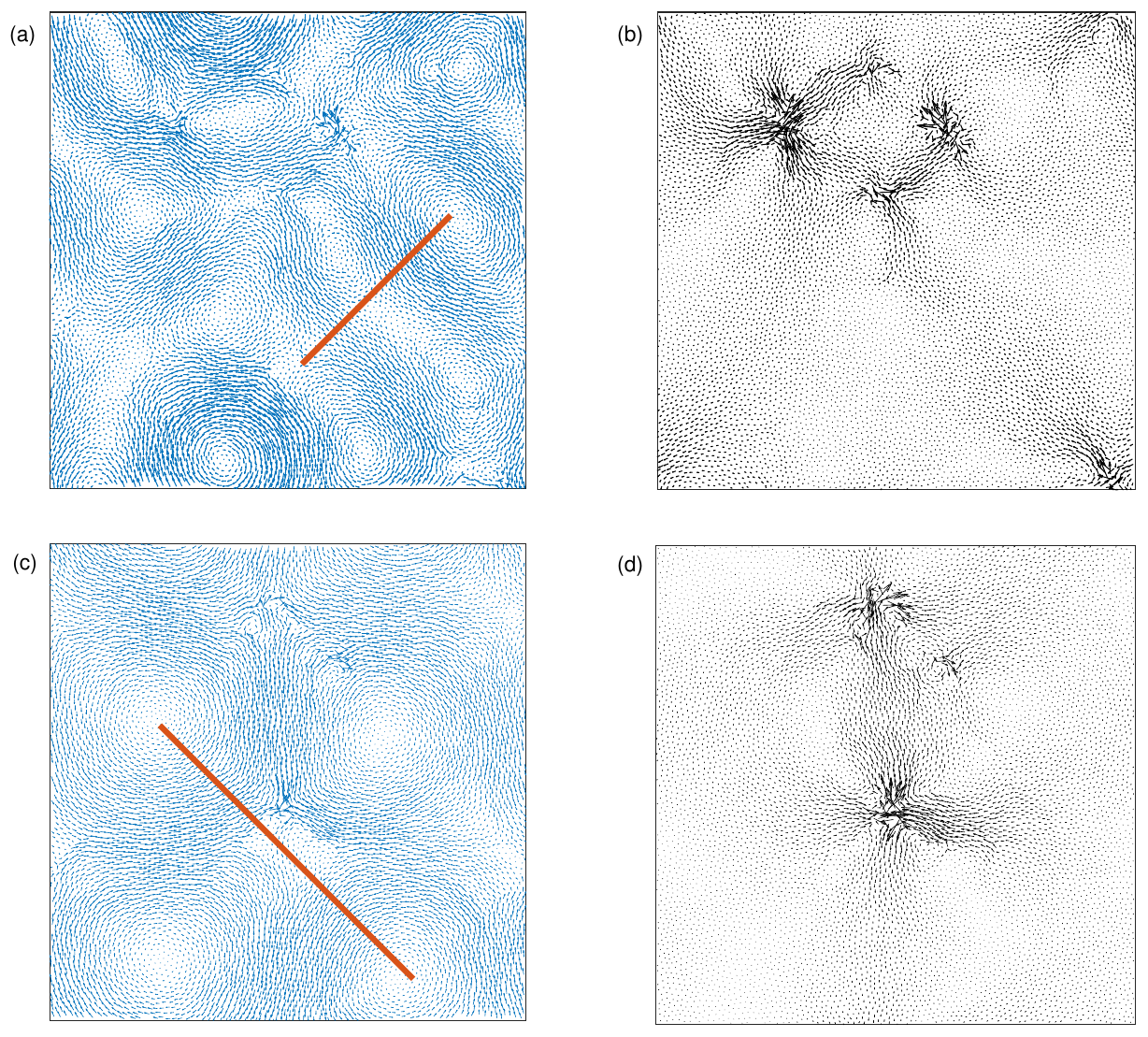}
  \caption{\footnotesize Comparison between vibrational modes $\psiv$ with (panel (a)) $\omega\!\approx\!\wbp/2$ and $\omega\!\approx\!\wbp/4$ (panel (c)), and the corresponding constructed superpositions $\phiv_\dagger$ (panels (b) and (d), respectively) of quasilocalized excitations extracted as explained in Sect.~\ref{sec:superposition} of the main text. The superpositions show that the ``imperfections'' seen in the wave patterns of these lower frequency vibrational modes in fact originate from hybridization with quasilocalized excitations. The red bars represent the phononic wavelength $2\pi c_{\rm s}/\omega$.}
  \label{fig:fig5}
\end{figure*}

In addition, future work should establish the generality of the conclusions arrived at in this work, across different glass-forming models, across different glass-formation protocols, and also in three dimensions, and make them more quantitative. In particular, it would be interesting to examine whether the removal of internal stresses in model disordered solids --- as done e.g.~in Refs.~\cite{eric_boson_peak_emt,inst_note,jcp_letter_scattering_2021} ---, or the introduction of strong attractive forces --- as done e.g.~in Ref.~\cite{sticky_spheres1_karina_pre2021} --- have a measurable effect on the nature of vibrational modes at the boson-peak frequency. 

Finally, we remark that methods to exhaustively compute the entire population of QLEs from a single computer glass are still under developement; once such tools become available, future research should quantify the physics of QLE-QLE and QLE-phonon hybridizations. Such studies will inform us about how vibrational modes are expected to behave in glasses approaching the thermodynamic $N\!\to\!\infty$ limit. In this context, we note that the computational results presented in this contribution has very recently led to some closely related theoretical and experimental (reanalysis of available data) progress~\cite{moriel2023boson}. 

\vspace{-0.2cm}

\acknowledgements
\vspace{-0.2cm}
We thank Karina Gonz\'alez-L\'opez and David Richard for their comments on the manuscript. E.L.~acknowledges Support from the NWO (Vidi grant no.~680-47-554/3259). E.B.~acknowledges support from the Ben May Center for Chemical Theory and Computation and the Harold Perlman Family.

\vspace{-0.2cm}

\appendix

\section{Lower-frequency vibrations}
\label{sec:AppendixA}

\begin{figure}[ht!]
  \includegraphics[width = 0.5\textwidth]{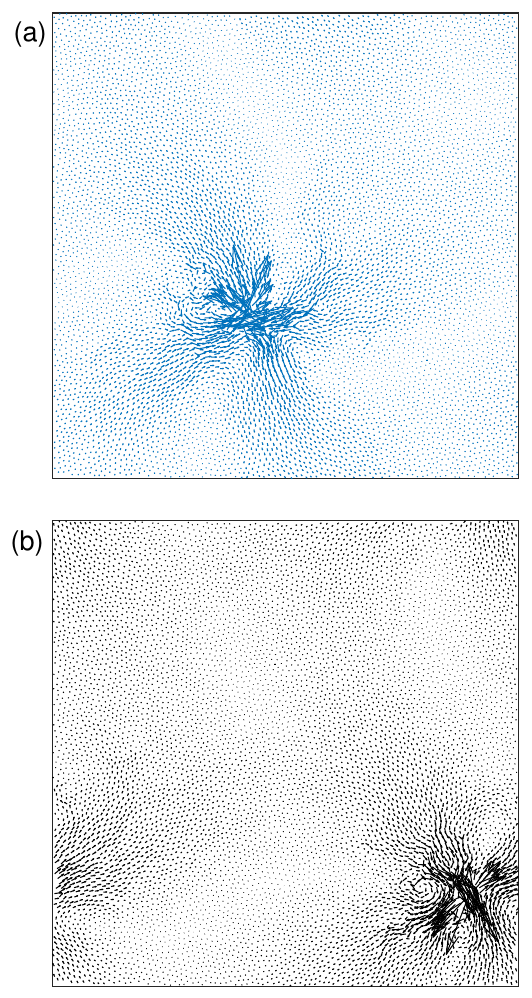}
  \caption{\footnotesize Examples of detected QLEs. The frequencies of these modes are $\omega\!=\!0.69$ (panel (a)) and $\omega\!=\!1.78$ (panel (b)). In both cases, the detected QLEs reveal the generic structure of a disordered core decorated by quadrupolar, algebraically decaying far-fields.}
  \label{fig:fig6}
\end{figure}

In Fig.~\ref{fig:fig5}, we show for comparison the results of the same analysis performed in Fig.~\ref{fig:fig3} of the main text, this time applied to a vibrational mode with frequency $\approx\wbp/4$ (top row), and to a vibrational mode with frequency $\approx\wbp/2$ (bottom row). It is observed that the ``imperfections'' seen in the wave patterns of vibrational modes at these lower frequency in fact originate from hybridization with quasilocalized excitations.

\section{Detected QLEs}
\label{sec:AppendixB}

In Fig.~\ref{fig:fig6}, we show the highest and lowest frequency QLEs detected using the method explained in Sect.~\ref{sec:qlm_detection}. The QLEs shown here have $\omega\!=\!0.69$ (panel (a)) and $\omega\!=\!1.78$ (panel (b)). Both modes show the same generic structural features of a disordered core decorated by quadrupolar algebraic far-field decays.

\newpage


\begin{thebibliography}{46}%
\makeatletter
\providecommand \@ifxundefined [1]{%
 \@ifx{#1\undefined}
}%
\providecommand \@ifnum [1]{%
 \ifnum #1\expandafter \@firstoftwo
 \else \expandafter \@secondoftwo
 \fi
}%
\providecommand \@ifx [1]{%
 \ifx #1\expandafter \@firstoftwo
 \else \expandafter \@secondoftwo
 \fi
}%
\providecommand \natexlab [1]{#1}%
\providecommand \enquote  [1]{``#1''}%
\providecommand \bibnamefont  [1]{#1}%
\providecommand \bibfnamefont [1]{#1}%
\providecommand \citenamefont [1]{#1}%
\providecommand \href@noop [0]{\@secondoftwo}%
\providecommand \href [0]{\begingroup \@sanitize@url \@href}%
\providecommand \@href[1]{\@@startlink{#1}\@@href}%
\providecommand \@@href[1]{\endgroup#1\@@endlink}%
\providecommand \@sanitize@url [0]{\catcode `\\12\catcode `\$12\catcode
  `\&12\catcode `\#12\catcode `\^12\catcode `\_12\catcode `\%12\relax}%
\providecommand \@@startlink[1]{}%
\providecommand \@@endlink[0]{}%
\providecommand \url  [0]{\begingroup\@sanitize@url \@url }%
\providecommand \@url [1]{\endgroup\@href {#1}{\urlprefix }}%
\providecommand \urlprefix  [0]{URL }%
\providecommand \Eprint [0]{\href }%
\providecommand \doibase [0]{https://doi.org/}%
\providecommand \selectlanguage [0]{\@gobble}%
\providecommand \bibinfo  [0]{\@secondoftwo}%
\providecommand \bibfield  [0]{\@secondoftwo}%
\providecommand \translation [1]{[#1]}%
\providecommand \BibitemOpen [0]{}%
\providecommand \bibitemStop [0]{}%
\providecommand \bibitemNoStop [0]{.\EOS\space}%
\providecommand \EOS [0]{\spacefactor3000\relax}%
\providecommand \BibitemShut  [1]{\csname bibitem#1\endcsname}%
\let\auto@bib@innerbib\@empty
\bibitem [{\citenamefont {Kittel}(2005)}]{kittel2005introduction}%
  \BibitemOpen
  \bibfield  {author} {\bibinfo {author} {\bibfnamefont {C.}~\bibnamefont
  {Kittel}},\ }\href@noop {} {\emph {\bibinfo {title} {Introduction to solid
  state physics}}}\ (\bibinfo  {publisher} {Wiley},\ \bibinfo {year}
  {2005})\BibitemShut {NoStop}%
\bibitem [{\citenamefont {Xu}\ \emph {et~al.}(2009)\citenamefont {Xu},
  \citenamefont {Vitelli}, \citenamefont {Wyart}, \citenamefont {Liu},\ and\
  \citenamefont {Nagel}}]{vincenzo_transport_prl_2009}%
  \BibitemOpen
  \bibfield  {author} {\bibinfo {author} {\bibfnamefont {N.}~\bibnamefont
  {Xu}}, \bibinfo {author} {\bibfnamefont {V.}~\bibnamefont {Vitelli}},
  \bibinfo {author} {\bibfnamefont {M.}~\bibnamefont {Wyart}}, \bibinfo
  {author} {\bibfnamefont {A.~J.}\ \bibnamefont {Liu}},\ and\ \bibinfo {author}
  {\bibfnamefont {S.~R.}\ \bibnamefont {Nagel}},\ }\bibfield  {title} {\bibinfo
  {title} {Energy transport in jammed sphere packings},\ }\href
  {https://doi.org/10.1103/PhysRevLett.102.038001} {\bibfield  {journal}
  {\bibinfo  {journal} {Phys. Rev. Lett.}\ }\textbf {\bibinfo {volume} {102}},\
  \bibinfo {pages} {038001} (\bibinfo {year} {2009})}\BibitemShut {NoStop}%
\bibitem [{\citenamefont {Lutsko}(1989)}]{lutsko}%
  \BibitemOpen
  \bibfield  {author} {\bibinfo {author} {\bibfnamefont {J.~F.}\ \bibnamefont
  {Lutsko}},\ }\bibfield  {title} {\bibinfo {title} {Generalized expressions
  for the calculation of elastic constants by computer simulation},\ }\href
  {https://doi.org/10.1063/1.342716} {\bibfield  {journal} {\bibinfo  {journal}
  {J. Appl. Phys.}\ }\textbf {\bibinfo {volume} {65}},\ \bibinfo {pages} {2991}
  (\bibinfo {year} {1989})}\BibitemShut {NoStop}%
\bibitem [{\citenamefont {Hentschel}\ \emph {et~al.}(2011)\citenamefont
  {Hentschel}, \citenamefont {Karmakar}, \citenamefont {Lerner},\ and\
  \citenamefont {Procaccia}}]{exist}%
  \BibitemOpen
  \bibfield  {author} {\bibinfo {author} {\bibfnamefont {H.~G.~E.}\
  \bibnamefont {Hentschel}}, \bibinfo {author} {\bibfnamefont {S.}~\bibnamefont
  {Karmakar}}, \bibinfo {author} {\bibfnamefont {E.}~\bibnamefont {Lerner}},\
  and\ \bibinfo {author} {\bibfnamefont {I.}~\bibnamefont {Procaccia}},\
  }\bibfield  {title} {\bibinfo {title} {Do athermal amorphous solids exist?},\
  }\href {https://doi.org/10.1103/PhysRevE.83.061101} {\bibfield  {journal}
  {\bibinfo  {journal} {Phys. Rev. E}\ }\textbf {\bibinfo {volume} {83}},\
  \bibinfo {pages} {061101} (\bibinfo {year} {2011})}\BibitemShut {NoStop}%
\bibitem [{\citenamefont {Richard}\ \emph
  {et~al.}(2020{\natexlab{a}})\citenamefont {Richard}, \citenamefont {Ozawa},
  \citenamefont {Patinet}, \citenamefont {Stanifer}, \citenamefont {Shang},
  \citenamefont {Ridout}, \citenamefont {Xu}, \citenamefont {Zhang},
  \citenamefont {Morse}, \citenamefont {Barrat}, \citenamefont {Berthier},
  \citenamefont {Falk}, \citenamefont {Guan}, \citenamefont {Liu},
  \citenamefont {Martens}, \citenamefont {Sastry}, \citenamefont
  {Vandembroucq}, \citenamefont {Lerner},\ and\ \citenamefont
  {Manning}}]{david_collaboration_2020}%
  \BibitemOpen
  \bibfield  {author} {\bibinfo {author} {\bibfnamefont {D.}~\bibnamefont
  {Richard}}, \bibinfo {author} {\bibfnamefont {M.}~\bibnamefont {Ozawa}},
  \bibinfo {author} {\bibfnamefont {S.}~\bibnamefont {Patinet}}, \bibinfo
  {author} {\bibfnamefont {E.}~\bibnamefont {Stanifer}}, \bibinfo {author}
  {\bibfnamefont {B.}~\bibnamefont {Shang}}, \bibinfo {author} {\bibfnamefont
  {S.~A.}\ \bibnamefont {Ridout}}, \bibinfo {author} {\bibfnamefont
  {B.}~\bibnamefont {Xu}}, \bibinfo {author} {\bibfnamefont {G.}~\bibnamefont
  {Zhang}}, \bibinfo {author} {\bibfnamefont {P.~K.}\ \bibnamefont {Morse}},
  \bibinfo {author} {\bibfnamefont {J.-L.}\ \bibnamefont {Barrat}}, \bibinfo
  {author} {\bibfnamefont {L.}~\bibnamefont {Berthier}}, \bibinfo {author}
  {\bibfnamefont {M.~L.}\ \bibnamefont {Falk}}, \bibinfo {author}
  {\bibfnamefont {P.}~\bibnamefont {Guan}}, \bibinfo {author} {\bibfnamefont
  {A.~J.}\ \bibnamefont {Liu}}, \bibinfo {author} {\bibfnamefont
  {K.}~\bibnamefont {Martens}}, \bibinfo {author} {\bibfnamefont
  {S.}~\bibnamefont {Sastry}}, \bibinfo {author} {\bibfnamefont
  {D.}~\bibnamefont {Vandembroucq}}, \bibinfo {author} {\bibfnamefont
  {E.}~\bibnamefont {Lerner}},\ and\ \bibinfo {author} {\bibfnamefont {M.~L.}\
  \bibnamefont {Manning}},\ }\bibfield  {title} {\bibinfo {title} {Predicting
  plasticity in disordered solids from structural indicators},\ }\href
  {https://doi.org/10.1103/PhysRevMaterials.4.113609} {\bibfield  {journal}
  {\bibinfo  {journal} {Phys. Rev. Materials}\ }\textbf {\bibinfo {volume}
  {4}},\ \bibinfo {pages} {113609} (\bibinfo {year}
  {2020}{\natexlab{a}})}\BibitemShut {NoStop}%
\bibitem [{\citenamefont {Richard}\ \emph
  {et~al.}(2021{\natexlab{a}})\citenamefont {Richard}, \citenamefont {Lerner},\
  and\ \citenamefont {Bouchbinder}}]{david_fracture_mrs_2021}%
  \BibitemOpen
  \bibfield  {author} {\bibinfo {author} {\bibfnamefont {D.}~\bibnamefont
  {Richard}}, \bibinfo {author} {\bibfnamefont {E.}~\bibnamefont {Lerner}},\
  and\ \bibinfo {author} {\bibfnamefont {E.}~\bibnamefont {Bouchbinder}},\
  }\bibfield  {title} {\bibinfo {title} {Brittle-to-ductile transitions in
  glasses: Roles of soft defects and loading geometry},\ }\href
  {https://doi.org/10.1557/s43577-021-00171-8} {\bibfield  {journal} {\bibinfo
  {journal} {MRS Bulletin}\ }\textbf {\bibinfo {volume} {46}},\ \bibinfo
  {pages} {902} (\bibinfo {year} {2021}{\natexlab{a}})}\BibitemShut {NoStop}%
\bibitem [{\citenamefont {Buchenau}\ \emph {et~al.}(1991)\citenamefont
  {Buchenau}, \citenamefont {Galperin}, \citenamefont {Gurevich},\ and\
  \citenamefont {Schober}}]{soft_potential_model_1991}%
  \BibitemOpen
  \bibfield  {author} {\bibinfo {author} {\bibfnamefont {U.}~\bibnamefont
  {Buchenau}}, \bibinfo {author} {\bibfnamefont {Y.~M.}\ \bibnamefont
  {Galperin}}, \bibinfo {author} {\bibfnamefont {V.~L.}\ \bibnamefont
  {Gurevich}},\ and\ \bibinfo {author} {\bibfnamefont {H.~R.}\ \bibnamefont
  {Schober}},\ }\bibfield  {title} {\bibinfo {title} {Anharmonic potentials and
  vibrational localization in glasses},\ }\href
  {https://doi.org/10.1103/PhysRevB.43.5039} {\bibfield  {journal} {\bibinfo
  {journal} {Phys. Rev. B}\ }\textbf {\bibinfo {volume} {43}},\ \bibinfo
  {pages} {5039} (\bibinfo {year} {1991})}\BibitemShut {NoStop}%
\bibitem [{\citenamefont {Schober}\ and\ \citenamefont
  {Oligschleger}(1996)}]{Schober_Oligschleger_numerics_PRB}%
  \BibitemOpen
  \bibfield  {author} {\bibinfo {author} {\bibfnamefont {H.~R.}\ \bibnamefont
  {Schober}}\ and\ \bibinfo {author} {\bibfnamefont {C.}~\bibnamefont
  {Oligschleger}},\ }\bibfield  {title} {\bibinfo {title} {Low-frequency
  vibrations in a model glass},\ }\href
  {https://doi.org/10.1103/PhysRevB.53.11469} {\bibfield  {journal} {\bibinfo
  {journal} {Phys. Rev. B}\ }\textbf {\bibinfo {volume} {53}},\ \bibinfo
  {pages} {11469} (\bibinfo {year} {1996})}\BibitemShut {NoStop}%
\bibitem [{\citenamefont {Gurevich}\ \emph {et~al.}(2003)\citenamefont
  {Gurevich}, \citenamefont {Parshin},\ and\ \citenamefont
  {Schober}}]{Gurevich2003}%
  \BibitemOpen
  \bibfield  {author} {\bibinfo {author} {\bibfnamefont {V.~L.}\ \bibnamefont
  {Gurevich}}, \bibinfo {author} {\bibfnamefont {D.~A.}\ \bibnamefont
  {Parshin}},\ and\ \bibinfo {author} {\bibfnamefont {H.~R.}\ \bibnamefont
  {Schober}},\ }\bibfield  {title} {\bibinfo {title} {Anharmonicity,
  vibrational instability, and the boson peak in glasses},\ }\href
  {https://doi.org/10.1103/PhysRevB.67.094203} {\bibfield  {journal} {\bibinfo
  {journal} {Phys. Rev. B}\ }\textbf {\bibinfo {volume} {67}},\ \bibinfo
  {pages} {094203} (\bibinfo {year} {2003})}\BibitemShut {NoStop}%
\bibitem [{\citenamefont {Parshin}\ \emph {et~al.}(2007)\citenamefont
  {Parshin}, \citenamefont {Schober},\ and\ \citenamefont
  {Gurevich}}]{Gurevich2007}%
  \BibitemOpen
  \bibfield  {author} {\bibinfo {author} {\bibfnamefont {D.~A.}\ \bibnamefont
  {Parshin}}, \bibinfo {author} {\bibfnamefont {H.~R.}\ \bibnamefont
  {Schober}},\ and\ \bibinfo {author} {\bibfnamefont {V.~L.}\ \bibnamefont
  {Gurevich}},\ }\bibfield  {title} {\bibinfo {title} {Vibrational instability,
  two-level systems, and the boson peak in glasses},\ }\href
  {https://doi.org/10.1103/PhysRevB.76.064206} {\bibfield  {journal} {\bibinfo
  {journal} {Phys. Rev. B}\ }\textbf {\bibinfo {volume} {76}},\ \bibinfo
  {pages} {064206} (\bibinfo {year} {2007})}\BibitemShut {NoStop}%
\bibitem [{\citenamefont {Lubchenko}\ and\ \citenamefont
  {Wolynes}(2003)}]{Lubchenko1515}%
  \BibitemOpen
  \bibfield  {author} {\bibinfo {author} {\bibfnamefont {V.}~\bibnamefont
  {Lubchenko}}\ and\ \bibinfo {author} {\bibfnamefont {P.~G.}\ \bibnamefont
  {Wolynes}},\ }\bibfield  {title} {\bibinfo {title} {The origin of the boson
  peak and thermal conductivity plateau in low-temperature glasses},\ }\href
  {https://doi.org/10.1073/pnas.252786999} {\bibfield  {journal} {\bibinfo
  {journal} {Proc. Natl. Acad. Sci. U.S.A.}\ }\textbf {\bibinfo {volume}
  {100}},\ \bibinfo {pages} {1515} (\bibinfo {year} {2003})}\BibitemShut
  {NoStop}%
\bibitem [{\citenamefont {Wyart}\ \emph {et~al.}(2005)\citenamefont {Wyart},
  \citenamefont {Silbert}, \citenamefont {Nagel},\ and\ \citenamefont
  {Witten}}]{matthieu_PRE_2005}%
  \BibitemOpen
  \bibfield  {author} {\bibinfo {author} {\bibfnamefont {M.}~\bibnamefont
  {Wyart}}, \bibinfo {author} {\bibfnamefont {L.~E.}\ \bibnamefont {Silbert}},
  \bibinfo {author} {\bibfnamefont {S.~R.}\ \bibnamefont {Nagel}},\ and\
  \bibinfo {author} {\bibfnamefont {T.~A.}\ \bibnamefont {Witten}},\ }\bibfield
   {title} {\bibinfo {title} {Effects of compression on the vibrational modes
  of marginally jammed solids},\ }\href
  {https://doi.org/10.1103/PhysRevE.72.051306} {\bibfield  {journal} {\bibinfo
  {journal} {Phys. Rev. E}\ }\textbf {\bibinfo {volume} {72}},\ \bibinfo
  {pages} {051306} (\bibinfo {year} {2005})}\BibitemShut {NoStop}%
\bibitem [{\citenamefont {Shintani}\ and\ \citenamefont
  {Tanaka}(2008)}]{tanaka_boson_peak_2008}%
  \BibitemOpen
  \bibfield  {author} {\bibinfo {author} {\bibfnamefont {H.}~\bibnamefont
  {Shintani}}\ and\ \bibinfo {author} {\bibfnamefont {H.}~\bibnamefont
  {Tanaka}},\ }\bibfield  {title} {\bibinfo {title} {Universal link between the
  boson peak and transverse phonons in glass},\ }\href
  {https://doi.org/10.1038/nmat2293} {\bibfield  {journal} {\bibinfo  {journal}
  {Nat. Mater.}\ }\textbf {\bibinfo {volume} {7}},\ \bibinfo {pages} {870}
  (\bibinfo {year} {2008})}\BibitemShut {NoStop}%
\bibitem [{\citenamefont {Ilyin}\ \emph {et~al.}(2009)\citenamefont {Ilyin},
  \citenamefont {Procaccia}, \citenamefont {Regev},\ and\ \citenamefont
  {Shokef}}]{procaccia_boson_peak_prb_2009}%
  \BibitemOpen
  \bibfield  {author} {\bibinfo {author} {\bibfnamefont {V.}~\bibnamefont
  {Ilyin}}, \bibinfo {author} {\bibfnamefont {I.}~\bibnamefont {Procaccia}},
  \bibinfo {author} {\bibfnamefont {I.}~\bibnamefont {Regev}},\ and\ \bibinfo
  {author} {\bibfnamefont {Y.}~\bibnamefont {Shokef}},\ }\bibfield  {title}
  {\bibinfo {title} {Randomness-induced redistribution of vibrational
  frequencies in amorphous solids},\ }\href
  {https://doi.org/10.1103/PhysRevB.80.174201} {\bibfield  {journal} {\bibinfo
  {journal} {Phys. Rev. B}\ }\textbf {\bibinfo {volume} {80}},\ \bibinfo
  {pages} {174201} (\bibinfo {year} {2009})}\BibitemShut {NoStop}%
\bibitem [{\citenamefont {Chumakov}\ \emph {et~al.}(2011)\citenamefont
  {Chumakov}, \citenamefont {Monaco}, \citenamefont {Monaco}, \citenamefont
  {Crichton}, \citenamefont {Bosak}, \citenamefont {R\"uffer}, \citenamefont
  {Meyer}, \citenamefont {Kargl}, \citenamefont {Comez}, \citenamefont
  {Fioretto}, \citenamefont {Giefers}, \citenamefont {Roitsch}, \citenamefont
  {Wortmann}, \citenamefont {Manghnani}, \citenamefont {Hushur}, \citenamefont
  {Williams}, \citenamefont {Balogh}, \citenamefont
  {Parli\ifmmode~\acute{n}\else \'{n}\fi{}ski}, \citenamefont {Jochym},\ and\
  \citenamefont {Piekarz}}]{Chumakov_2011_bosonPeak}%
  \BibitemOpen
  \bibfield  {author} {\bibinfo {author} {\bibfnamefont {A.~I.}\ \bibnamefont
  {Chumakov}}, \bibinfo {author} {\bibfnamefont {G.}~\bibnamefont {Monaco}},
  \bibinfo {author} {\bibfnamefont {A.}~\bibnamefont {Monaco}}, \bibinfo
  {author} {\bibfnamefont {W.~A.}\ \bibnamefont {Crichton}}, \bibinfo {author}
  {\bibfnamefont {A.}~\bibnamefont {Bosak}}, \bibinfo {author} {\bibfnamefont
  {R.}~\bibnamefont {R\"uffer}}, \bibinfo {author} {\bibfnamefont
  {A.}~\bibnamefont {Meyer}}, \bibinfo {author} {\bibfnamefont
  {F.}~\bibnamefont {Kargl}}, \bibinfo {author} {\bibfnamefont
  {L.}~\bibnamefont {Comez}}, \bibinfo {author} {\bibfnamefont
  {D.}~\bibnamefont {Fioretto}}, \bibinfo {author} {\bibfnamefont
  {H.}~\bibnamefont {Giefers}}, \bibinfo {author} {\bibfnamefont
  {S.}~\bibnamefont {Roitsch}}, \bibinfo {author} {\bibfnamefont
  {G.}~\bibnamefont {Wortmann}}, \bibinfo {author} {\bibfnamefont {M.~H.}\
  \bibnamefont {Manghnani}}, \bibinfo {author} {\bibfnamefont {A.}~\bibnamefont
  {Hushur}}, \bibinfo {author} {\bibfnamefont {Q.}~\bibnamefont {Williams}},
  \bibinfo {author} {\bibfnamefont {J.}~\bibnamefont {Balogh}}, \bibinfo
  {author} {\bibfnamefont {K.}~\bibnamefont {Parli\ifmmode~\acute{n}\else
  \'{n}\fi{}ski}}, \bibinfo {author} {\bibfnamefont {P.}~\bibnamefont
  {Jochym}},\ and\ \bibinfo {author} {\bibfnamefont {P.}~\bibnamefont
  {Piekarz}},\ }\bibfield  {title} {\bibinfo {title} {Equivalence of the boson
  peak in glasses to the transverse acoustic van hove singularity in
  crystals},\ }\href {https://doi.org/10.1103/PhysRevLett.106.225501}
  {\bibfield  {journal} {\bibinfo  {journal} {Phys. Rev. Lett.}\ }\textbf
  {\bibinfo {volume} {106}},\ \bibinfo {pages} {225501} (\bibinfo {year}
  {2011})}\BibitemShut {NoStop}%
\bibitem [{\citenamefont {Hong}\ \emph {et~al.}(2011)\citenamefont {Hong},
  \citenamefont {Novikov},\ and\ \citenamefont
  {Sokolov}}]{sokolov_boson_peak_scale}%
  \BibitemOpen
  \bibfield  {author} {\bibinfo {author} {\bibfnamefont {L.}~\bibnamefont
  {Hong}}, \bibinfo {author} {\bibfnamefont {V.~N.}\ \bibnamefont {Novikov}},\
  and\ \bibinfo {author} {\bibfnamefont {A.~P.}\ \bibnamefont {Sokolov}},\
  }\bibfield  {title} {\bibinfo {title} {Dynamic heterogeneities, boson peak,
  and activation volume in glass-forming liquids},\ }\href
  {https://doi.org/10.1103/PhysRevE.83.061508} {\bibfield  {journal} {\bibinfo
  {journal} {Phys. Rev. E}\ }\textbf {\bibinfo {volume} {83}},\ \bibinfo
  {pages} {061508} (\bibinfo {year} {2011})}\BibitemShut {NoStop}%
\bibitem [{\citenamefont {Marruzzo}\ \emph {et~al.}(2013)\citenamefont
  {Marruzzo}, \citenamefont {Schirmacher}, \citenamefont {Fratalocchi},\ and\
  \citenamefont {Ruocco}}]{Schirmacher_2013_boson_peak}%
  \BibitemOpen
  \bibfield  {author} {\bibinfo {author} {\bibfnamefont {A.}~\bibnamefont
  {Marruzzo}}, \bibinfo {author} {\bibfnamefont {W.}~\bibnamefont
  {Schirmacher}}, \bibinfo {author} {\bibfnamefont {A.}~\bibnamefont
  {Fratalocchi}},\ and\ \bibinfo {author} {\bibfnamefont {G.}~\bibnamefont
  {Ruocco}},\ }\bibfield  {title} {\bibinfo {title} {Heterogeneous shear
  elasticity of glasses: the origin of the boson peak},\ }\href
  {https://doi.org/10.1038/srep01407} {\bibfield  {journal} {\bibinfo
  {journal} {Sci. Rep.}\ }\textbf {\bibinfo {volume} {3}},\ \bibinfo {pages}
  {1407 EP } (\bibinfo {year} {2013})}\BibitemShut {NoStop}%
\bibitem [{\citenamefont {DeGiuli}\ \emph {et~al.}(2014)\citenamefont
  {DeGiuli}, \citenamefont {Laversanne-Finot}, \citenamefont {During},
  \citenamefont {Lerner},\ and\ \citenamefont {Wyart}}]{eric_boson_peak_emt}%
  \BibitemOpen
  \bibfield  {author} {\bibinfo {author} {\bibfnamefont {E.}~\bibnamefont
  {DeGiuli}}, \bibinfo {author} {\bibfnamefont {A.}~\bibnamefont
  {Laversanne-Finot}}, \bibinfo {author} {\bibfnamefont {G.}~\bibnamefont
  {During}}, \bibinfo {author} {\bibfnamefont {E.}~\bibnamefont {Lerner}},\
  and\ \bibinfo {author} {\bibfnamefont {M.}~\bibnamefont {Wyart}},\ }\bibfield
   {title} {\bibinfo {title} {Effects of coordination and pressure on sound
  attenuation{,} boson peak and elasticity in amorphous solids},\ }\href
  {https://doi.org/10.1039/C4SM00561A} {\bibfield  {journal} {\bibinfo
  {journal} {Soft Matter}\ }\textbf {\bibinfo {volume} {10}},\ \bibinfo {pages}
  {5628} (\bibinfo {year} {2014})}\BibitemShut {NoStop}%
\bibitem [{\citenamefont {Franz}\ \emph {et~al.}(2015)\citenamefont {Franz},
  \citenamefont {Parisi}, \citenamefont {Urbani},\ and\ \citenamefont
  {Zamponi}}]{silvio}%
  \BibitemOpen
  \bibfield  {author} {\bibinfo {author} {\bibfnamefont {S.}~\bibnamefont
  {Franz}}, \bibinfo {author} {\bibfnamefont {G.}~\bibnamefont {Parisi}},
  \bibinfo {author} {\bibfnamefont {P.}~\bibnamefont {Urbani}},\ and\ \bibinfo
  {author} {\bibfnamefont {F.}~\bibnamefont {Zamponi}},\ }\bibfield  {title}
  {\bibinfo {title} {Universal spectrum of normal modes in low-temperature
  glasses},\ }\href {https://doi.org/10.1073/pnas.1511134112} {\bibfield
  {journal} {\bibinfo  {journal} {Proc. Natl. Acad. Sci. U.S.A.}\ }\textbf
  {\bibinfo {volume} {112}},\ \bibinfo {pages} {14539} (\bibinfo {year}
  {2015})}\BibitemShut {NoStop}%
\bibitem [{\citenamefont {Lerner}\ and\ \citenamefont
  {Bouchbinder}(2021)}]{JCP_Perspective}%
  \BibitemOpen
  \bibfield  {author} {\bibinfo {author} {\bibfnamefont {E.}~\bibnamefont
  {Lerner}}\ and\ \bibinfo {author} {\bibfnamefont {E.}~\bibnamefont
  {Bouchbinder}},\ }\bibfield  {title} {\bibinfo {title} {Low-energy
  quasilocalized excitations in structural glasses},\ }\href
  {https://doi.org/10.1063/5.0069477} {\bibfield  {journal} {\bibinfo
  {journal} {J. Chem. Phys.}\ }\textbf {\bibinfo {volume} {155}},\ \bibinfo
  {pages} {200901} (\bibinfo {year} {2021})}\BibitemShut {NoStop}%
\bibitem [{\citenamefont {Hu}\ and\ \citenamefont
  {Tanaka}(2022)}]{tanaka_2d_modes_2022}%
  \BibitemOpen
  \bibfield  {author} {\bibinfo {author} {\bibfnamefont {Y.-C.}\ \bibnamefont
  {Hu}}\ and\ \bibinfo {author} {\bibfnamefont {H.}~\bibnamefont {Tanaka}},\
  }\bibfield  {title} {\bibinfo {title} {Origin of the boson peak in amorphous
  solids},\ }\href {https://doi.org/10.1038/s41567-022-01628-6} {\bibfield
  {journal} {\bibinfo  {journal} {Nat. Phys.}\ }\textbf {\bibinfo {volume}
  {18}},\ \bibinfo {pages} {669} (\bibinfo {year} {2022})}\BibitemShut
  {NoStop}%
\bibitem [{\citenamefont {Lerner}\ and\ \citenamefont
  {Bouchbinder}(2022)}]{disordered_crystals_prl_2022}%
  \BibitemOpen
  \bibfield  {author} {\bibinfo {author} {\bibfnamefont {E.}~\bibnamefont
  {Lerner}}\ and\ \bibinfo {author} {\bibfnamefont {E.}~\bibnamefont
  {Bouchbinder}},\ }\bibfield  {title} {\bibinfo {title} {Disordered crystals
  reveal soft quasilocalized glassy excitations},\ }\href
  {https://doi.org/10.1103/PhysRevLett.129.095501} {\bibfield  {journal}
  {\bibinfo  {journal} {Phys. Rev. Lett.}\ }\textbf {\bibinfo {volume} {129}},\
  \bibinfo {pages} {095501} (\bibinfo {year} {2022})}\BibitemShut {NoStop}%
\bibitem [{\citenamefont {Alexander}(1998)}]{shlomo}%
  \BibitemOpen
  \bibfield  {author} {\bibinfo {author} {\bibfnamefont {S.}~\bibnamefont
  {Alexander}},\ }\bibfield  {title} {\bibinfo {title} {Amorphous solids: their
  structure, lattice dynamics and elasticity},\ }\href
  {https://doi.org/http://dx.doi.org/10.1016/S0370-1573(97)00069-0} {\bibfield
  {journal} {\bibinfo  {journal} {Phys. Rep.}\ }\textbf {\bibinfo {volume}
  {296}},\ \bibinfo {pages} {65 } (\bibinfo {year} {1998})}\BibitemShut
  {NoStop}%
\bibitem [{\citenamefont {Baldi}\ \emph {et~al.}(2010)\citenamefont {Baldi},
  \citenamefont {Giordano}, \citenamefont {Monaco},\ and\ \citenamefont
  {Ruta}}]{baldi2010sound}%
  \BibitemOpen
  \bibfield  {author} {\bibinfo {author} {\bibfnamefont {G.}~\bibnamefont
  {Baldi}}, \bibinfo {author} {\bibfnamefont {V.~M.}\ \bibnamefont {Giordano}},
  \bibinfo {author} {\bibfnamefont {G.}~\bibnamefont {Monaco}},\ and\ \bibinfo
  {author} {\bibfnamefont {B.}~\bibnamefont {Ruta}},\ }\bibfield  {title}
  {\bibinfo {title} {Sound attenuation at terahertz frequencies and the boson
  peak of vitreous silica},\ }\href
  {https://doi.org/10.1103/PhysRevLett.104.195501} {\bibfield  {journal}
  {\bibinfo  {journal} {Phys. Rev. Lett.}\ }\textbf {\bibinfo {volume} {104}},\
  \bibinfo {pages} {195501} (\bibinfo {year} {2010})}\BibitemShut {NoStop}%
\bibitem [{\citenamefont {Wyart}(2010)}]{mw_EM_epl}%
  \BibitemOpen
  \bibfield  {author} {\bibinfo {author} {\bibfnamefont {M.}~\bibnamefont
  {Wyart}},\ }\bibfield  {title} {\bibinfo {title} {Scaling of phononic
  transport with connectivity in amorphous solids},\ }\href
  {http://stacks.iop.org/0295-5075/89/i=6/a=64001} {\bibfield  {journal}
  {\bibinfo  {journal} {Europhys. Lett.}\ }\textbf {\bibinfo {volume} {89}},\
  \bibinfo {pages} {64001} (\bibinfo {year} {2010})}\BibitemShut {NoStop}%
\bibitem [{\citenamefont {Yan}\ \emph {et~al.}(2016)\citenamefont {Yan},
  \citenamefont {DeGiuli},\ and\ \citenamefont
  {Wyart}}]{new_variational_argument_epl_2016}%
  \BibitemOpen
  \bibfield  {author} {\bibinfo {author} {\bibfnamefont {L.}~\bibnamefont
  {Yan}}, \bibinfo {author} {\bibfnamefont {E.}~\bibnamefont {DeGiuli}},\ and\
  \bibinfo {author} {\bibfnamefont {M.}~\bibnamefont {Wyart}},\ }\bibfield
  {title} {\bibinfo {title} {On variational arguments for vibrational modes
  near jamming},\ }\href {http://stacks.iop.org/0295-5075/114/i=2/a=26003}
  {\bibfield  {journal} {\bibinfo  {journal} {Europhys. Lett.}\ }\textbf
  {\bibinfo {volume} {114}},\ \bibinfo {pages} {26003} (\bibinfo {year}
  {2016})}\BibitemShut {NoStop}%
\bibitem [{\citenamefont {Duval}\ and\ \citenamefont
  {Mermet}(1998)}]{Duval_prb_1998}%
  \BibitemOpen
  \bibfield  {author} {\bibinfo {author} {\bibfnamefont {E.}~\bibnamefont
  {Duval}}\ and\ \bibinfo {author} {\bibfnamefont {A.}~\bibnamefont {Mermet}},\
  }\bibfield  {title} {\bibinfo {title} {Inelastic x-ray scattering from
  nonpropagating vibrational modes in glasses},\ }\href
  {https://doi.org/10.1103/PhysRevB.58.8159} {\bibfield  {journal} {\bibinfo
  {journal} {Phys. Rev. B}\ }\textbf {\bibinfo {volume} {58}},\ \bibinfo
  {pages} {8159} (\bibinfo {year} {1998})}\BibitemShut {NoStop}%
\bibitem [{\citenamefont {Taraskin}\ \emph {et~al.}(2001)\citenamefont
  {Taraskin}, \citenamefont {Loh}, \citenamefont {Natarajan},\ and\
  \citenamefont {Elliott}}]{taraskin_lattice_disorder_prl_2001}%
  \BibitemOpen
  \bibfield  {author} {\bibinfo {author} {\bibfnamefont {S.~N.}\ \bibnamefont
  {Taraskin}}, \bibinfo {author} {\bibfnamefont {Y.~L.}\ \bibnamefont {Loh}},
  \bibinfo {author} {\bibfnamefont {G.}~\bibnamefont {Natarajan}},\ and\
  \bibinfo {author} {\bibfnamefont {S.~R.}\ \bibnamefont {Elliott}},\
  }\bibfield  {title} {\bibinfo {title} {Origin of the boson peak in systems
  with lattice disorder},\ }\href {https://doi.org/10.1103/PhysRevLett.86.1255}
  {\bibfield  {journal} {\bibinfo  {journal} {Phys. Rev. Lett.}\ }\textbf
  {\bibinfo {volume} {86}},\ \bibinfo {pages} {1255} (\bibinfo {year}
  {2001})}\BibitemShut {NoStop}%
\bibitem [{\citenamefont {Parisi}(2003)}]{Parisi_boson_peak_2003}%
  \BibitemOpen
  \bibfield  {author} {\bibinfo {author} {\bibfnamefont {G.}~\bibnamefont
  {Parisi}},\ }\bibfield  {title} {\bibinfo {title} {On the origin of the boson
  peak},\ }\href {https://doi.org/10.1088/0953-8984/15/11/302} {\bibfield
  {journal} {\bibinfo  {journal} {J. Phys. Condens. Matter}\ }\textbf {\bibinfo
  {volume} {15}},\ \bibinfo {pages} {S765} (\bibinfo {year}
  {2003})}\BibitemShut {NoStop}%
\bibitem [{\citenamefont {Lerner}\ \emph {et~al.}(2016)\citenamefont {Lerner},
  \citenamefont {D\"uring},\ and\ \citenamefont
  {Bouchbinder}}]{modes_prl_2016}%
  \BibitemOpen
  \bibfield  {author} {\bibinfo {author} {\bibfnamefont {E.}~\bibnamefont
  {Lerner}}, \bibinfo {author} {\bibfnamefont {G.}~\bibnamefont {D\"uring}},\
  and\ \bibinfo {author} {\bibfnamefont {E.}~\bibnamefont {Bouchbinder}},\
  }\bibfield  {title} {\bibinfo {title} {Statistics and properties of
  low-frequency vibrational modes in structural glasses},\ }\href
  {https://doi.org/10.1103/PhysRevLett.117.035501} {\bibfield  {journal}
  {\bibinfo  {journal} {Phys. Rev. Lett.}\ }\textbf {\bibinfo {volume} {117}},\
  \bibinfo {pages} {035501} (\bibinfo {year} {2016})}\BibitemShut {NoStop}%
\bibitem [{\citenamefont {Richard}\ \emph
  {et~al.}(2020{\natexlab{b}})\citenamefont {Richard}, \citenamefont
  {Gonz\'alez-L\'opez}, \citenamefont {Kapteijns}, \citenamefont {Pater},
  \citenamefont {Vaknin}, \citenamefont {Bouchbinder},\ and\ \citenamefont
  {Lerner}}]{modes_prl_2020}%
  \BibitemOpen
  \bibfield  {author} {\bibinfo {author} {\bibfnamefont {D.}~\bibnamefont
  {Richard}}, \bibinfo {author} {\bibfnamefont {K.}~\bibnamefont
  {Gonz\'alez-L\'opez}}, \bibinfo {author} {\bibfnamefont {G.}~\bibnamefont
  {Kapteijns}}, \bibinfo {author} {\bibfnamefont {R.}~\bibnamefont {Pater}},
  \bibinfo {author} {\bibfnamefont {T.}~\bibnamefont {Vaknin}}, \bibinfo
  {author} {\bibfnamefont {E.}~\bibnamefont {Bouchbinder}},\ and\ \bibinfo
  {author} {\bibfnamefont {E.}~\bibnamefont {Lerner}},\ }\bibfield  {title}
  {\bibinfo {title} {Universality of the nonphononic vibrational spectrum
  across different classes of computer glasses},\ }\href
  {https://doi.org/10.1103/PhysRevLett.125.085502} {\bibfield  {journal}
  {\bibinfo  {journal} {Phys. Rev. Lett.}\ }\textbf {\bibinfo {volume} {125}},\
  \bibinfo {pages} {085502} (\bibinfo {year} {2020}{\natexlab{b}})}\BibitemShut
  {NoStop}%
\bibitem [{\citenamefont {Rainone}\ \emph {et~al.}(2020)\citenamefont
  {Rainone}, \citenamefont {Bouchbinder},\ and\ \citenamefont
  {Lerner}}]{pinching_pnas}%
  \BibitemOpen
  \bibfield  {author} {\bibinfo {author} {\bibfnamefont {C.}~\bibnamefont
  {Rainone}}, \bibinfo {author} {\bibfnamefont {E.}~\bibnamefont
  {Bouchbinder}},\ and\ \bibinfo {author} {\bibfnamefont {E.}~\bibnamefont
  {Lerner}},\ }\bibfield  {title} {\bibinfo {title} {Pinching a glass reveals
  key properties of its soft spots},\ }\href
  {https://doi.org/10.1073/pnas.1919958117} {\bibfield  {journal} {\bibinfo
  {journal} {Proc. Natl. Acad. Sci. U.S.A.}\ }\textbf {\bibinfo {volume}
  {117}},\ \bibinfo {pages} {5228} (\bibinfo {year} {2020})}\BibitemShut
  {NoStop}%
\bibitem [{\citenamefont {Wang}\ \emph {et~al.}(2019)\citenamefont {Wang},
  \citenamefont {Ninarello}, \citenamefont {Guan}, \citenamefont {Berthier},
  \citenamefont {Szamel},\ and\ \citenamefont {Flenner}}]{LB_modes_2019}%
  \BibitemOpen
  \bibfield  {author} {\bibinfo {author} {\bibfnamefont {L.}~\bibnamefont
  {Wang}}, \bibinfo {author} {\bibfnamefont {A.}~\bibnamefont {Ninarello}},
  \bibinfo {author} {\bibfnamefont {P.}~\bibnamefont {Guan}}, \bibinfo {author}
  {\bibfnamefont {L.}~\bibnamefont {Berthier}}, \bibinfo {author}
  {\bibfnamefont {G.}~\bibnamefont {Szamel}},\ and\ \bibinfo {author}
  {\bibfnamefont {E.}~\bibnamefont {Flenner}},\ }\bibfield  {title} {\bibinfo
  {title} {Low-frequency vibrational modes of stable glasses},\ }\href
  {https://doi.org/10.1038/s41467-018-07978-1} {\bibfield  {journal} {\bibinfo
  {journal} {Nat. Commun.}\ }\textbf {\bibinfo {volume} {10}},\ \bibinfo
  {pages} {26} (\bibinfo {year} {2019})}\BibitemShut {NoStop}%
\bibitem [{\citenamefont {Bouchbinder}\ and\ \citenamefont
  {Lerner}(2018)}]{phonon_widths}%
  \BibitemOpen
  \bibfield  {author} {\bibinfo {author} {\bibfnamefont {E.}~\bibnamefont
  {Bouchbinder}}\ and\ \bibinfo {author} {\bibfnamefont {E.}~\bibnamefont
  {Lerner}},\ }\bibfield  {title} {\bibinfo {title} {Universal disorder-induced
  broadening of phonon bands: from disordered lattices to glasses},\ }\href
  {http://stacks.iop.org/1367-2630/20/i=7/a=073022} {\bibfield  {journal}
  {\bibinfo  {journal} {New J. Phys.}\ }\textbf {\bibinfo {volume} {20}},\
  \bibinfo {pages} {073022} (\bibinfo {year} {2018})}\BibitemShut {NoStop}%
\bibitem [{\citenamefont {Kapteijns}\ \emph {et~al.}(2018)\citenamefont
  {Kapteijns}, \citenamefont {Bouchbinder},\ and\ \citenamefont
  {Lerner}}]{modes_prl_2018}%
  \BibitemOpen
  \bibfield  {author} {\bibinfo {author} {\bibfnamefont {G.}~\bibnamefont
  {Kapteijns}}, \bibinfo {author} {\bibfnamefont {E.}~\bibnamefont
  {Bouchbinder}},\ and\ \bibinfo {author} {\bibfnamefont {E.}~\bibnamefont
  {Lerner}},\ }\bibfield  {title} {\bibinfo {title} {Universal nonphononic
  density of states in 2d, 3d, and 4d glasses},\ }\href
  {https://doi.org/10.1103/PhysRevLett.121.055501} {\bibfield  {journal}
  {\bibinfo  {journal} {Phys. Rev. Lett.}\ }\textbf {\bibinfo {volume} {121}},\
  \bibinfo {pages} {055501} (\bibinfo {year} {2018})}\BibitemShut {NoStop}%
\bibitem [{\citenamefont {Richard}\ \emph
  {et~al.}(2021{\natexlab{b}})\citenamefont {Richard}, \citenamefont
  {Kapteijns}, \citenamefont {Giannini}, \citenamefont {Manning},\ and\
  \citenamefont {Lerner}}]{pseudo_harmonic_prl}%
  \BibitemOpen
  \bibfield  {author} {\bibinfo {author} {\bibfnamefont {D.}~\bibnamefont
  {Richard}}, \bibinfo {author} {\bibfnamefont {G.}~\bibnamefont {Kapteijns}},
  \bibinfo {author} {\bibfnamefont {J.~A.}\ \bibnamefont {Giannini}}, \bibinfo
  {author} {\bibfnamefont {M.~L.}\ \bibnamefont {Manning}},\ and\ \bibinfo
  {author} {\bibfnamefont {E.}~\bibnamefont {Lerner}},\ }\bibfield  {title}
  {\bibinfo {title} {Simple and broadly applicable definition of shear
  transformation zones},\ }\href
  {https://doi.org/10.1103/PhysRevLett.126.015501} {\bibfield  {journal}
  {\bibinfo  {journal} {Phys. Rev. Lett.}\ }\textbf {\bibinfo {volume} {126}},\
  \bibinfo {pages} {015501} (\bibinfo {year} {2021}{\natexlab{b}})}\BibitemShut
  {NoStop}%
\bibitem [{\citenamefont {Lerner}\ and\ \citenamefont
  {Bouchbinder}(2018{\natexlab{a}})}]{cge_paper}%
  \BibitemOpen
  \bibfield  {author} {\bibinfo {author} {\bibfnamefont {E.}~\bibnamefont
  {Lerner}}\ and\ \bibinfo {author} {\bibfnamefont {E.}~\bibnamefont
  {Bouchbinder}},\ }\bibfield  {title} {\bibinfo {title} {A characteristic
  energy scale in glasses},\ }\href {https://doi.org/10.1063/1.5024776}
  {\bibfield  {journal} {\bibinfo  {journal} {J. Chem. Phys.}\ }\textbf
  {\bibinfo {volume} {148}},\ \bibinfo {pages} {214502} (\bibinfo {year}
  {2018}{\natexlab{a}})}\BibitemShut {NoStop}%
\bibitem [{\citenamefont {Gonz\'alez-L\'opez}\ \emph
  {et~al.}(2021)\citenamefont {Gonz\'alez-L\'opez}, \citenamefont {Shivam},
  \citenamefont {Zheng}, \citenamefont {Ciamarra},\ and\ \citenamefont
  {Lerner}}]{sticky_spheres1_karina_pre2021}%
  \BibitemOpen
  \bibfield  {author} {\bibinfo {author} {\bibfnamefont {K.}~\bibnamefont
  {Gonz\'alez-L\'opez}}, \bibinfo {author} {\bibfnamefont {M.}~\bibnamefont
  {Shivam}}, \bibinfo {author} {\bibfnamefont {Y.}~\bibnamefont {Zheng}},
  \bibinfo {author} {\bibfnamefont {M.~P.}\ \bibnamefont {Ciamarra}},\ and\
  \bibinfo {author} {\bibfnamefont {E.}~\bibnamefont {Lerner}},\ }\bibfield
  {title} {\bibinfo {title} {Mechanical disorder of sticky-sphere glasses. i.
  effect of attractive interactions},\ }\href
  {https://doi.org/10.1103/PhysRevE.103.022605} {\bibfield  {journal} {\bibinfo
   {journal} {Phys. Rev. E}\ }\textbf {\bibinfo {volume} {103}},\ \bibinfo
  {pages} {022605} (\bibinfo {year} {2021})}\BibitemShut {NoStop}%
\bibitem [{\citenamefont {Gartner}\ and\ \citenamefont
  {Lerner}(2016)}]{SciPost2016}%
  \BibitemOpen
  \bibfield  {author} {\bibinfo {author} {\bibfnamefont {L.}~\bibnamefont
  {Gartner}}\ and\ \bibinfo {author} {\bibfnamefont {E.}~\bibnamefont
  {Lerner}},\ }\bibfield  {title} {\bibinfo {title} {{Nonlinear modes
  disentangle glassy and Goldstone modes in structural glasses}},\ }\href
  {https://doi.org/10.21468/SciPostPhys.1.2.016} {\bibfield  {journal}
  {\bibinfo  {journal} {SciPost Phys.}\ }\textbf {\bibinfo {volume} {1}},\
  \bibinfo {pages} {016} (\bibinfo {year} {2016})}\BibitemShut {NoStop}%
\bibitem [{\citenamefont {Kapteijns}\ \emph {et~al.}(2020)\citenamefont
  {Kapteijns}, \citenamefont {Richard},\ and\ \citenamefont
  {Lerner}}]{episode_1_2020}%
  \BibitemOpen
  \bibfield  {author} {\bibinfo {author} {\bibfnamefont {G.}~\bibnamefont
  {Kapteijns}}, \bibinfo {author} {\bibfnamefont {D.}~\bibnamefont {Richard}},\
  and\ \bibinfo {author} {\bibfnamefont {E.}~\bibnamefont {Lerner}},\
  }\bibfield  {title} {\bibinfo {title} {Nonlinear quasilocalized excitations
  in glasses: True representatives of soft spots},\ }\href
  {https://doi.org/10.1103/PhysRevE.101.032130} {\bibfield  {journal} {\bibinfo
   {journal} {Phys. Rev. E}\ }\textbf {\bibinfo {volume} {101}},\ \bibinfo
  {pages} {032130} (\bibinfo {year} {2020})}\BibitemShut {NoStop}%
\bibitem [{pac()}]{package}%
  \BibitemOpen
  \bibinfo {note} {Softspot PYTHON package (2020),
  \href{https://pypi.org/project/softspot/}{https://pypi.org/project/softspot/}.}\BibitemShut
  {Stop}%
\bibitem [{foo()}]{footnote}%
  \BibitemOpen
  \bibinfo {note} {Given a QLM $\piv$, we define its frequency as
  $\omega\!\equiv\!\sqrt{\piv\cdot\calBold{H}\cdot\piv}$.}\BibitemShut {Stop}%
\bibitem [{\citenamefont {Moriel}\ \emph {et~al.}(2023)\citenamefont {Moriel},
  \citenamefont {Lerner},\ and\ \citenamefont {Bouchbinder}}]{moriel2023boson}%
  \BibitemOpen
  \bibfield  {author} {\bibinfo {author} {\bibfnamefont {A.}~\bibnamefont
  {Moriel}}, \bibinfo {author} {\bibfnamefont {E.}~\bibnamefont {Lerner}},\
  and\ \bibinfo {author} {\bibfnamefont {E.}~\bibnamefont {Bouchbinder}},\
  }\bibfield  {title} {\bibinfo {title} {The boson peak in the vibrational
  spectra of glasses},\ }\href {https://doi.org/10.48550/arXiv.2304.03661}
  {\bibfield  {journal} {\bibinfo  {journal} {arXiv preprint arXiv:2304.03661}\
  } (\bibinfo {year} {2023})}\BibitemShut {NoStop}%
\bibitem [{\citenamefont {Vainer}\ \emph {et~al.}(2006)\citenamefont {Vainer},
  \citenamefont {Naumov}, \citenamefont {Bauer},\ and\ \citenamefont
  {Kador}}]{experimental_local_vibrations_boson_peak_prl_2006}%
  \BibitemOpen
  \bibfield  {author} {\bibinfo {author} {\bibfnamefont {Y.~G.}\ \bibnamefont
  {Vainer}}, \bibinfo {author} {\bibfnamefont {A.~V.}\ \bibnamefont {Naumov}},
  \bibinfo {author} {\bibfnamefont {M.}~\bibnamefont {Bauer}},\ and\ \bibinfo
  {author} {\bibfnamefont {L.}~\bibnamefont {Kador}},\ }\bibfield  {title}
  {\bibinfo {title} {Experimental evidence of the local character of vibrations
  constituting the boson peak in amorphous solids},\ }\href
  {https://doi.org/10.1103/PhysRevLett.97.185501} {\bibfield  {journal}
  {\bibinfo  {journal} {Phys. Rev. Lett.}\ }\textbf {\bibinfo {volume} {97}},\
  \bibinfo {pages} {185501} (\bibinfo {year} {2006})}\BibitemShut {NoStop}%
\bibitem [{\citenamefont {Lerner}\ and\ \citenamefont
  {Bouchbinder}(2018{\natexlab{b}})}]{inst_note}%
  \BibitemOpen
  \bibfield  {author} {\bibinfo {author} {\bibfnamefont {E.}~\bibnamefont
  {Lerner}}\ and\ \bibinfo {author} {\bibfnamefont {E.}~\bibnamefont
  {Bouchbinder}},\ }\bibfield  {title} {\bibinfo {title} {Frustration-induced
  internal stresses are responsible for quasilocalized modes in structural
  glasses},\ }\href {https://doi.org/10.1103/PhysRevE.97.032140} {\bibfield
  {journal} {\bibinfo  {journal} {Phys. Rev. E}\ }\textbf {\bibinfo {volume}
  {97}},\ \bibinfo {pages} {032140} (\bibinfo {year}
  {2018}{\natexlab{b}})}\BibitemShut {NoStop}%
\bibitem [{\citenamefont {Kapteijns}\ \emph {et~al.}(2021)\citenamefont
  {Kapteijns}, \citenamefont {Richard}, \citenamefont {Bouchbinder},\ and\
  \citenamefont {Lerner}}]{jcp_letter_scattering_2021}%
  \BibitemOpen
  \bibfield  {author} {\bibinfo {author} {\bibfnamefont {G.}~\bibnamefont
  {Kapteijns}}, \bibinfo {author} {\bibfnamefont {D.}~\bibnamefont {Richard}},
  \bibinfo {author} {\bibfnamefont {E.}~\bibnamefont {Bouchbinder}},\ and\
  \bibinfo {author} {\bibfnamefont {E.}~\bibnamefont {Lerner}},\ }\bibfield
  {title} {\bibinfo {title} {Elastic moduli fluctuations predict wave
  attenuation rates in glasses},\ }\href {https://doi.org/10.1063/5.0038710}
  {\bibfield  {journal} {\bibinfo  {journal} {J. Chem. Phys.}\ }\textbf
  {\bibinfo {volume} {154}},\ \bibinfo {pages} {081101} (\bibinfo {year}
  {2021})}\BibitemShut {NoStop}%
\end{thebibliography}
%

\end{document}